# High-Power Testing of the APT Power Coupler

E. N. Schmierer, K. C. D. Chan, D. C. Gautier, J. G. Gioia, W. B. Haynes, F. L. Krawczyk, M. A. Madrid, D. L. Schrage, J. A. Waynert, LANL, Los Alamos, NM 87545, USA
B. Rusnak, LLNL, Livermore, CA 94550, USA

*Abstract*

For the baseline APT (Accelerator Production of Tritium) linac design, power couplers are required to transmit 210-kW of CW RF power to the superconducting cavities. These APT couplers operate at 700 MHz, have a coaxial design and an adjustable coupling to the superconducting cavities. Since May 1999, we have been testing couplers of this design on a room-temperature test stand. We completed tests at transmitted-power and reflected-power conditions up to 1 MW. We also tested the couplers with a portion of the outer conductor cooled by liquid nitrogen. Under this latter condition, we studied the effects of condensed gases on coupler performance. The results of these tests indicate that the APT couplers are capable of delivering more than 500 kW to the cavities. We are in the process of increasing the baseline coupler design requirement to 420 kW of transmitted power to take advantage of this successful development. In this paper, we describe the results of our high-power coupler tests.

## 1 INTRODUCTION

The APT power coupler is one of the challenges for the APT (Accelerator Production of Tritium) linac, which has a high-energy superconducting (SC) section spanning 210–1030 MeV [1]. Due to a high current of 100 mA, a large amount of CW RF power must be transmitted through these couplers to the SC cavities. In 1997, 180 kW was the highest CW power transmitted through couplers to a particle beam, which was limited by multipacting and cooling to remove RF losses. At that time, a coupler requirement of 210 kW (at 700 MHz) was chosen for the APT couplers as a reasonable extrapolation of the technology, and two couplers would be needed to feed each cavity operating at a modest gradient of 5 MV/m.

Figure 1 shows the APT coupler design. Five design features were particularly employed to enhance its capability to robustly transmit high CW RF power: (1) A coaxial-type RF window with two ceramic disks was chosen to minimize window failures and their impacts; (2) Warm windows were chosen and located away from line of sight of the beam to minimize multipacting; (3) A coaxial line with a large diameter was chosen to increase the power thresholds of the multipacting band; (4) A large vacuum port at the coupler was included to insure good vacuum at the coupler and window; and (5) Extensive RF and thermal simulations were performed to reduce RF loss and to optimize cooling.

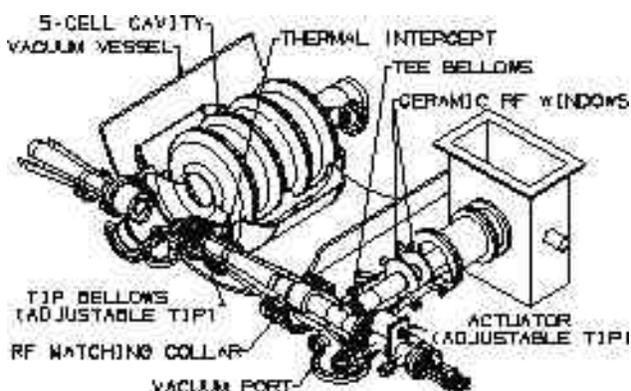

Figure 1. Schematic diagram of the APT power coupler

Details of the coupler design and results of low-power tests can be found in Ref. [2] and [3] respectively, and will not be repeated here. In this paper, we describe the high-power tests that we performed in the last twelve months.

## 2 HIGH-POWER TESTS

The high-power tests were performed on a specially designed test stand (Figure 2) [5]. The test stand was supplied with up to 1 MW of 700-MHz RF power from a klystron. RF power was transmitted from one coupler, through a copper cavity, to a second coupler, and then to a water-cooled RF load. The couplers and the copper cavity were well matched to provide 100% transmission. This arrangement allowed the testing of two couplers simultaneously. Since the test stand was at room temperature, the test conditions were slightly different from those conditions that would be if the couplers were in a cryomodule. Also, the inner conductors were cooled by air on the test stand instead of gaseous helium as is planned for the APT plant operation.

The test stand was equipped to perform three types of tests: (1) Transmitted-Power Capability; (2) Totally-Reflected Power; and (3) Condensed-Gas Effect. Details and results of these tests are described in Sections 2.1 to 2.3.

We tested couplers with fixed and adjustable couplings. The adjustability of a coupler is provided by tip bellows made of BeCu. Couplers with fixed coupling, which do

---

Work supported by US Department of energy

not have the tip bellows, are expected to offer more robust operation, but less flexibility during the APT linac operation where a range of beam currents will be accelerated.

Before testing, the components of the couplers were cleaned in a Class-1000 clean area. The outer conductors were rinsed with deionized-water (DI) and allowed to dry in the cleanroom. Using Gauge-32 copper wool, the inner conductors were scrubbed with DI water and methanol. Following scrubbing, they were rinsed with DI water and air-dried. The parts were assembled in the cleanroom and then transported to the coupler test stand. The components were inserted into the copper cavity and the RF window assemblies attached.

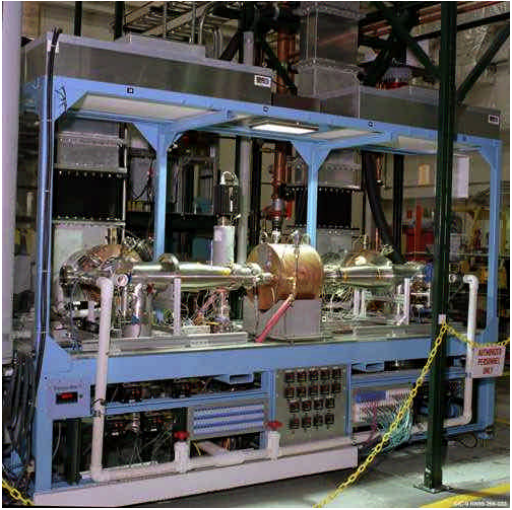

Figure 2. Test Stand for APT couplers

## 2.1 Transmitted-Power Capability Test

In the first test, RF power was transmitted through the couplers into the RF load. During conditioning, the power level was raised in steps to reach the highest power level possible. The size of the steps was defined by increases in vacuum level caused by increased heating and outgassing of the couplers. With a baseline vacuum of 8 x $10^{-9}$ Torr, the increase in vacuum was kept to below 1 x $10^{-6}$. During these tests, both fixed and adjustable couplers reached 1 MW of CW power.

## 2.2 Totally-Reflected Power Test

In the reflected-power test, the RF load was replaced with a "sliding short," which reflected all the forward power and set up a standing-wave pattern in the couplers. The total reflection of power was to simulate a situation during APT plant operation where the beam might be lost, reflecting nearly all the RF power. This test was conducted with the fixed couplers only. During testing, the maxima and minima of the standing wave were moved to five locations over a quarter wavelength along the system by adjusting the sliding-short position. For a sliding-short position corresponding to full current operation, the tests reached 950 kW CW. For other sliding-short positions, we reach a power level of 550 kW CW and a power level of 850 kW at a duty cycle of 50%.

## 2.3 Condensed Gas Test

Experiences of other laboratories have shown that gas condensation on power couplers enhances multipacting by changing the secondary emission coefficients of coupler surfaces and introducing vacuum bursts released by RF heating. To investigate the condensed-gas effects, we cooled the outer conductor by passing liquid nitrogen through the passages of the coupler heat exchanger originally designed for liquid helium. Figure 3 shows the measured temperature profiles at different power levels and the calculated temperature profile at 210 kW for the cryomodule and on test bed. Water vapor and $CO_2$ were condensed during these tests. Test results showed no enhancement of multipacting.

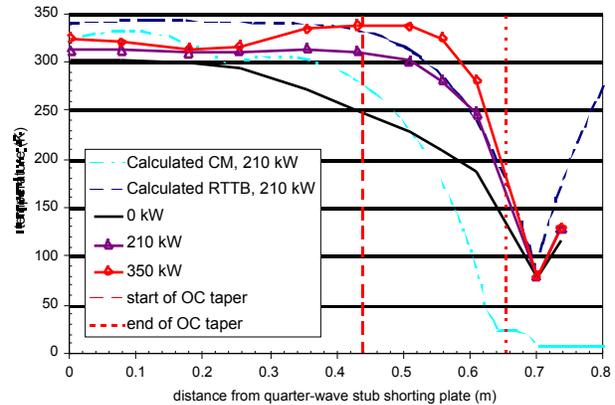

Figure 3. Cold test temperature profile of outer conductor

# 3 OBSERVATIONS DURING TESTS

In this section, we summarize our observations during the tests. These observations were interesting to high-power coupler operations. Their resolution will lead to future tests and design improvements.

## 3.1 Bellows Failures and Thermal Analysis

While testing power couplers with adjustable coupling, the BeCu tip bellows failed on two occasions after reaching a transmitted power level of 750 kW. Inspection of the failed tip bellows indicated that over-heating caused the failures. The proposed explanation of overheating was supported by thermal calculations. The calculations showed that because of the stagnation of cooling air in the convolutions and the lower thermal and RF electrical conductivities of BeCu (factors of four and two respectively compared to pure copper), the temperature of the tip bellows could be 800°C higher than the rest of the inner conductor. This is much higher than the design temperature and would lead to failure of BeCu material. Thermal simulations further showed that the temperature of the bellows could be effectively reduced by 400°C if we copper-plated the BeCu bellows and used helium gas as coolant (instead of air) during the test as is specified in the plant design.

### 3.2 Electron Etching

All power couplers tested exhibited evidence of electron activity on the surface of the inner conductors. Figure 4 shows a pattern of markings indicative of such activity. The patterns varied in concentration and position with test as well as with power level. At low power levels ( 500 kW), there were only shallow dendrical patterns. Above 500 kW, the patterns were deeper and more concentrated. Although the electron activity did not affect coupler operation, we are concerned with degradation of the components with time and deposition of copper from the inner conductor surface onto RF windows and cavities. We are planning additional tests to find the threshold of this "electron etching" and its dependence on other factors.

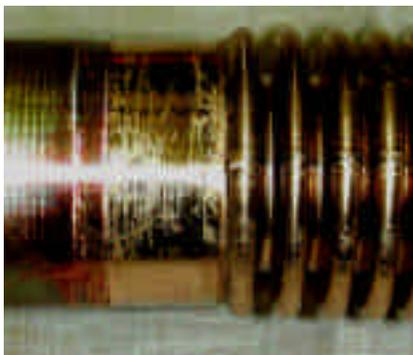

Figure 4. Markings indicative of electronic activities

### 3.3 Outgassing During Power Conditioning

During conditioning, partial pressures of residual gases were measured. Figure 5 shows these pressures as a function of conditioning time along with power levels during the testing of an adjustable coupler. Initially, these pressures increased with power level. At 500 kW, the pressures of $CO_2$ and water decreased while the pressure of hydrogen kept increasing.

The residual-gas pressures during cold tests are shown in Figure 6. As can be seen, the partial pressures increased starting from time 10:00 when RF power of 500 kW was applied. The partial pressures decreased starting at time 11:30 when the RF power level was reduced to 350 kW, and condensed residual gases were released starting at time 13:40 when liquid nitrogen ran out and the outer conductor warmed up.

## 4 SUMMARY AND CONCLUSION

We successfully tested the performance of the APT power couplers at high-power with respect to transmitted power capability, totally reflected power, and condensed-gas effects on multipacting. Data indicates that the APT power-coupler design is capable of handling power much higher than the APT requirement of 210 kW. We are in the process of changing the APT power-coupler requirement to 420 kW to reduce the number of couplers by a factor of two, which will lead to a reduction in the cost of the linac by $60M. We are presently conducting a second set of tests to insure that we can run the couplers robustly at a power level of 420 kW.

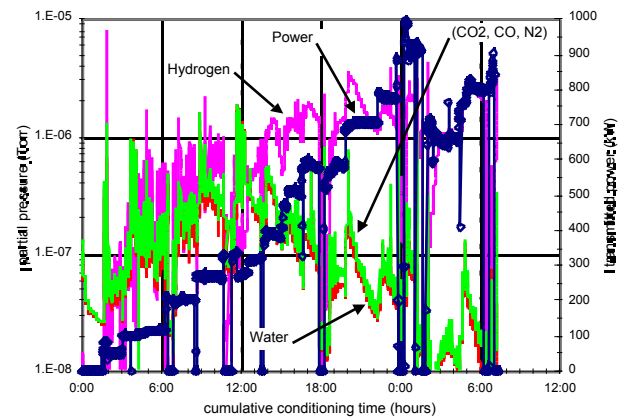

Figure 5. Residual gas vs. time compared with power level during conditioning

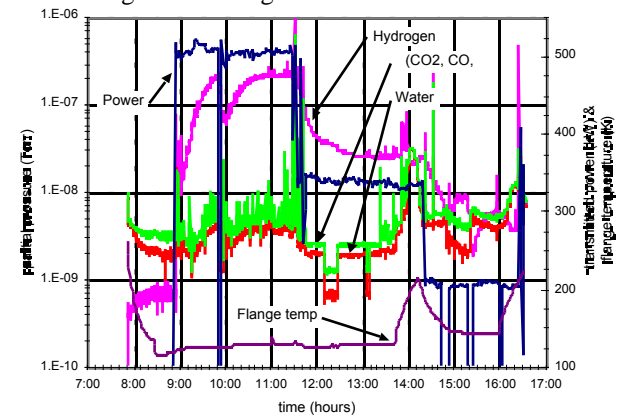

Figure 6. Residual gas vs. time compared with power level and outer conductor flange temperature during cold test